\documentclass{iopart}

\usepackage{graphicx}
\usepackage{fullpage}
\usepackage{latexsym}
\usepackage{amssymb}
\usepackage{wasysym}
\usepackage{color,linegoal}
\usepackage[procnames]{listings}

\definecolor{keywords}{RGB}{255,0,90}
\definecolor{comments}{RGB}{0,0,113}
\definecolor{red}{RGB}{160,0,0}
\definecolor{green}{RGB}{0,150,0}
\definecolor{mygray}{rgb}{0.5,0.5,0.5}

\newcommand{\um} {~\mu\mathrm{m}}
\newcommand{\rmdeg} {~\mathrm{deg}}

\newcommand{\taumin} {\tau_\mathrm{min}}
\newcommand{\tpp} {t_{pp}}

\begin{document}

\title{An efficient scheme for sampling fast dynamics at a low average data acquisition rate}

\author{A. Philippe$^{1}$, S. Aime$^{1}$, V. Roger$^{1}$, R. Jelinek$^{1}$, G. Pr\'{e}vot$^{1}$, L. Berthier$^{1}$, L. Cipelletti$^{1}$}
\address{$1$- Laboratoire Charles Coulomb (L2C), UMR 5221 CNRS-Universit\'{e} de Montpellier, Montpellier, F-France.}
\ead{luca.cipelletti@umontpellier.fr}


\begin{abstract}
We introduce a temporal scheme for data sampling, based on a variable delay between two successive data acquisitions. The scheme is designed so as to reduce the average data flow rate, while still retaining the information on the data evolution on fast time scales. The practical implementation of the scheme is discussed and demonstrated in light scattering and microscopy experiments that probe the dynamics of colloidal suspensions using CMOS or CCD cameras as detectors.
\end{abstract}

\maketitle

\section{Introduction}
\label{sec:intro}

Since the birth of modern science in the sixteenth century, measuring, quantifying and modelling how a system evolves in time has been one of the key challenges for physicists. For condensed matter systems comprising many particles, the time evolution is quantified by comparing system configurations at different times, or by studying the temporal fluctuations of a physical quantity directly related to the particle configuration. An example of the first approach is the particle mean squared displacement, which quantifies the average change of particle positions, as determined, e.g., in optical or confocal microscopy experiments with colloidal particles~\cite{Crocker1996,Elliot2001,Prasad2007}. The second method is exemplified by dynamic light scattering (DLS)~\cite{Berne1976}, which relates the temporal fluctuations of laser light scattered by the sample to its microscopic dynamics.

Both approaches require to sample repeatedly the system over time, which implies the acquisition of a stream of data. Modern scientific apparatuses often produce large amounts of data: this results in high-rate data flow, making data handling challenging. Two-dimensional (2D) detectors such as CMOS cameras illustrate nicely this challenge. Fast cameras that acquire images of several Mbytes at rates often exceeding 1 kHz are now affordable and increasingly popular in many setups, raising the issue of dealing with data flows of the order of Gbytes per second. Two-dimensional detectors are widely used in optical or confocal microscopy, e.g. in biology~\cite{Kherlopian2008}, in soft matter~\cite{Elliot2001,Prasad2007} or in microfluidics applications~\cite{Wereley2010}, but also in experiments based on conventional low-magnification imaging, e.g. for granular systems~\cite{Kudrolli2004} or in fluid dynamics~\cite{Adrian2005}. Moreover, two-dimensional detectors are increasingly replacing point-like detectors in techniques such as fluorescence imaging~\cite{Cubeddu2002} or in the multispeckle approach~\cite{Kirsch1996} to DLS and X-photon correlation spectroscopy~\cite{Leheny2012}. They are also at the heart of recently introduced techniques that combine features of scattering and imaging, such as photon correlation imaging~\cite{Duri2009,Cipelletti2013} or differential dynamic microscopy~\cite{Cerbino2008} and other digital Fourier microscopy techniques~\cite{Cerbino2014}.

In this paper, we describe a scheme for acquiring data at a low \emph{average} rate, while still preserving the information on the fast dynamics of the system. For the sake of concreteness, we will assume that the data are 2D images and illustrate the scheme with examples from scattering and microscopy experiments; however, we emphasize that the scheme is quite general and may be applied to the acquisition of any kind of data, possibly as a function of variables different from time (e.g. when sampling some sample property over space). Existing acquisition schemes typically consist in sampling the system at a constant rate or, in a more refined version, at a rate that slowly changes in time to adapt to a possible evolution of the system dynamics~\cite{Brunel2007}. The drawback of this approach is two-fold: firstly, if the dynamics of interest span several order of magnitudes or the system evolution has to be followed over a long time, a very large amount of data has to be acquired and processed. Secondly, the rate at which a detector can acquire data often exceeds the rate at which data can be processed or stored for later processing. This is typically the case of modern cameras, whose acquisition rate may exceed that at which images can be written to a hard disk (HD), sometimes even if state-of-the-art solid state devices or arrays of independent disks (RAID) are used. Under these conditions, one has to reduce the acquisition rate to match the processing or storage rate, thereby not fully exploiting the capabilities of the detector.

The multitau scheme, first proposed in traditional DLS~\cite{schatzel_singlephoton_1993} and later extended to multispeckle DLS~\cite{lucacipelletti_ultralowangle_1999} and microscopy-based microrheology measurements~\cite{yanagishima_realtime_2011,evans_direct_2009}, addresses these issues by coarse-graining the data over time. Several coarse-graining levels are implemented in parallel, allowing one to characterize the system evolution via temporal correlation functions (one per coarsening level) that span a large range of time delays with a limited number of channels. This method is particularly well-suited for processing the data on-the-fly, yielding low-noise correlation functions thanks to the massive averaging associated with coarse-graining. However, the rate at which data are acquired and processed decreases with increasing coarse-graining level. This makes it impossible to capture rapid fluctuations of the dynamics at large time delays, as observed, e.g., in the temporally heterogeneous dynamics of many glassy systems~\cite{berthier_dynamical_2011}. Additionally, the multitau scheme is based on fast, constant-rate data acquisition, which typically makes it impossible to write the data to the HD for later additional processing or for checking purposes. An alternative method could consist in alternating short bursts of fast acquisitions, where the images are transferred to a fast memory storage (e.g. the computer RAM or the on-board memory of the camera or the frame grabber), with long stretches of time where data are acquired at a lower rate and written to the HD. During these long stretches of time, the RAM data acquired in the previous burst should be copied to the HD. The main drawback of such a scheme is the uneven distribution of the fast and slow acquisition phases over time: if the system dynamics are not stationary (e.g. due to aging or dynamical heterogeneity~\cite{Cipelletti2005}), one misses all changes of the fast dynamics in between two burst phases.

The method introduced in this work addresses these challenges by using a variable-delay acquisition scheme. As it will be shown, the method deliberately under-samples the data with respect to the maximum rate allowed by the detector, so as to limit the data flow rate. However, the scheme is designed so as to interlace the fast and slow acquisition phases, so that the system dynamics is sampled as uniformly as possible in time. The paper is organized as follows: in Sec.~\ref{sec:theo} we introduce the new acquisition scheme and briefly discuss its practical implementation. Section~\ref{sec:mm} reviews the essential features of the DLS, DDM, and particle tracking methods and provides details on the experimental samples. 
The results of the light scattering and microscopy experiments are presented and discussed in Sec.~\ref{sec:results}, which is followed by some brief conclusions (Sec.~\ref{sec:conclusions}).

\section{Acquisition time scheme}
\label{sec:theo}
The acquisition scheme consists of a sequence of $2N$ images that is repeated cyclically. Each cycle is formed by two interlaced sub-sequences. The even images of the cycle are regularly spaced in time, every $\tpp$ seconds (see Fig.~1 
for an example with $\tpp = 1$ s). The index $pp$ stands for the time ``per pairs'' of images. The odd images are taken at a variable time delay $\tau_k$ with respect to the preceding even image. The time delay $\tau_k$ increases with $k$ as a power law, such that the $\tau_k$'s are regularly spaced in a logarithmic scale and cover the range between a minimum delay $\taumin$ and $\tpp$:
\begin{equation}
\tau_k = 10^{k/J}\taumin \,,
\label{eq:tauk}
\end{equation}
with $k=0,1,...,N-1$ and $J$ the desired number of time delays per decade. The total number of images per cycle is dictated by the ratio between the time per pair and the minimum delay, and by the number of sub-$\tpp$ time delays per decade. From Eq.~(\ref{eq:tauk}) and the constraint $\tau_k < \tpp$, one finds
\begin{equation}
N = \mathrm{ceil}\left( J\log_{10}\frac{\tpp}{\taumin}\right )\,,
\label{eq:N}
\end{equation}
where $\mathrm{ceil}(x)$ indicates the smallest integer $\ge x$. Each cycle comprises $2N$ images and lasts $N\tpp$; the acquisition times for the images belonging to the $M-th$ cycle are
\begin{eqnarray}
\label{eq:tm}
t_m = (M-1)N\tpp + \frac{m}{2}\tpp~&m = 0,2,4,...,2N-2\\
t_m= (M-1)N\tpp + \frac{m-1}{2}\tpp+10^{\frac{m-1}{2J}}\taumin~~~&m = 1,3,5,...,2N-1\,.
\end{eqnarray}
One may introduce a ``compression factor'' $\xi$ defined as the number of images that would have been acquired in a cycle with a traditional constant-delay scheme, divided by the number of images acquired over the same period with the variable-delay scheme, assuming the same minimum delay $\taumin$ in both cases. The compression factor is $\xi = (N\tpp\taumin^{-1})/(2N) = \tpp/2\taumin$, which can be of order 100 or larger.

As an illustration of the scheme, the bottom panel of Fig.~1 
shows the acquisition times for a cycle of $2N = 12$ images. The even images (open circles) are spaced by $\tpp = 1$s; the red crosses indicate the acquisition times for the odd images, each of which is delayed by $\tau$ with respect to the preceding image, with $\taumin = 0.015~\mathrm{s} \le \tau < \tpp$, and where $J=3$ logarithmically spaced sub-$\tpp$ delays per decade have been used (see top panel).

\begin{figure}[htbp]
\centering{
\includegraphics[width=8cm]{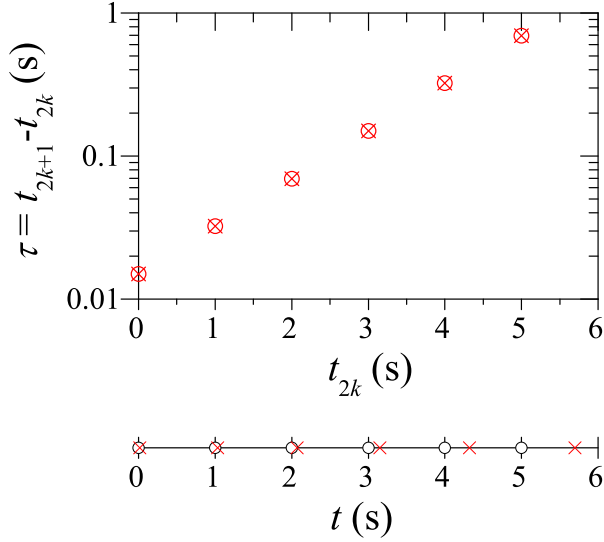}
\caption{Acquisition scheme for $\tpp = 1$ s, $J= 3$, $\taumin = 0.015$ s. For the sake of clarity, only the first cycle is shown. Bottom: acquisition times. The open circles correspond to the even images, spaced by $\tpp$, the red crosses to the odd images. The cycle contains a total of $2N = 12$ images. Top: time delay between an odd image and the preceding even image, as a function of the acquisition time of the even image.}}
\label{figscheme}
\end{figure}

Usually, $\taumin$ is chosen to be the smallest delay compatible with the camera specifications, i.e. $1/\taumin$ corresponds to the maximum frame rate. The \textit{average} acquisition rate, however, is $2/\tpp$, which can be set to be much lower than the maximum frame rate by choosing $\tpp >> \taumin$. This allows for enough time for the images to be, e.g., written to a hard disk or processed. In the following, we shall refer to any operation performed on the images after their acquisition as to `processing'. In order to decouple the acquisition process (which occurs at a time-varying rate, up to the maximum rate $1/\taumin$) from the image processing (which needs to be performed at a rate as uniform as possible, in order to cope with the hardware limitations), a buffering scheme must be used. As soon as they are acquired, the images are transferred to a buffer, whose memory space is physically located either in the PC RAM or on the frame grabber board, if available. This transfer is typically very fast and can be easily performed at an instantaneous rate equal to or even faster than the maximum camera frame rate. The buffer is read and emptied progressively by an image processing routine, at an instantaneous rate close to $2/\tpp$, the average data acquisition rate. In order to implement this buffering scheme, one should write a software with two separate yet synchronized threads, one for acquiring the images and one for processing them. In the experiments described below, we implement the buffering scheme in two different ways. For the light scattering experiments, the acquisition software is written in Labview FPGA, which has built-in routines for implementing the buffering scheme via a genuine multi-thread mechanism. For the microscopy experiments, we use a simple, single-thread software, where both the image acquisition and the image processing routines are called from the same loop, but the image processing routine is skipped when images have to be acquired rapidly (e.g. when the delay between consecutive images is equal to or slightly larger than $\taumin$), while it is called repeatedly to empty the buffer when enough time is available before the next image acquisition. A code snippet in Python illustrating this procedure is provided as Supplementary Data.

\section{Materials and methods}
\label{sec:mm}
\subsection{Multispeckle dynamic light scattering}
\label{sec:dls}
Dynamic light scattering~\cite{Berne1976} experiments are performed using a setup similar to that described in~\cite{ElMasri2005}. The sample is placed in a temperature-controlled copper holder and is illuminated by a laser beam with in-vacuo wavelength $\lambda = 532.5~\mathrm{nm}$. The scattered light is detected simultaneously by up to four CCD cameras (Pulnix TM-6740GE-w, images cropped to $640 \times 160$ pixels), placed at scattering angles in the range $15~\mathrm{deg} \leq \theta \leq 75~\mathrm{deg}$. For each CCD, the intensity correlation function $g_2(\tau)-1$ is calculated from a time series of images of the scattered light using the multispeckle~\cite{Kirsch1996} scheme:
\begin{equation}
\label{eq:multispeckle}
g_2(\tau)-1 = \left<c_I(t,\tau)\right>_t \,,
\end{equation}
where the time average is performed on the two-time degree of correlation~\cite{Duri2005}
\begin{equation}
\label{eq:cI}
c_I(t,\tau)= \frac{\left<I_p(t)I_p(t+\tau)\right>_p}{\left<I_p(t)\right>_p\left<I_p(t+\tau)\right>_p}-1 \,.
\end{equation}
Here, $I_p(t)$ is the intensity of the $p$-th pixel at time $t$ and $\left<\cdot \cdot \cdot\right>_p$ indicates an average over all CCD pixels, which are associated to a small solid angle centered around $\theta$. The purpose of the time average of Eq.~(\ref{eq:multispeckle}) is to reduce the experimental noise; it is performed over the full duration of the experiment for stationary samples, or over a short time window of duration $t_{exp}$ for samples whose dynamics evolve in time. When averaging over time, care has to be taken in order to extract from the variable-delay image sequence the appropriate pairs of images separated by a given time lag. The software provided as Supplemental Data illustrates how this can be accomplished.
The images are acquired and saved to hard disks using the scheme of Sec.~\ref{sec:theo}; they are then processed off-line to calculate $g_2-1$ according to Eqs.~(\ref{eq:multispeckle}) and (\ref{eq:cI}), correcting for the CCD electronic noise and the uneven sample illumination as detailed in~\cite{Duri2005}. The CCD cameras are triggered by a TTL signal, issued from a PICDEM 2 Plus card (by Microchip Technology Inc.) programmed using in-house C code, or by a National Instrument CompactRIO-9076 card with two TTL output C Ni-9402 modules, controlled via a custom Labview FPGA code.

Dynamic light scattering data are analyzed using the usual DLS formalism for Brownian particles. For a suspension of identical, non-interacting spherical particles, $g_2-1$ decays at a rate dictated by the particle diffusion coefficient $D$ and the scattering vector $q$~\cite{Berne1976}:
\begin{equation}
\label{eq:g2-1}
g_2(\tau)-1 = \exp\left ( -2q^2D\tau\right) \,,
\end{equation}
with $q = 4\pi n \lambda^{-1}\sin \theta/2$ and $n$ the solvent refractive index. In Sec.~\ref{sec:results} we will present data for melamine particles with diameter $2a = 1.14\um$ (Microparticles GmbH), suspended at a volume fraction $\varphi = 6\times 10^{-5}$ in a 2/98 w/w water/glycerol mixture, with viscosity $\eta = 290~\mathrm{mPa~s}$ at temperature $T = 20~^\circ \mathrm{C}$, and for a suspension of PNiPAM microgels, synthesized following~\cite{Senff1999}, for which the volume fraction, as calculated according to the definition of~\cite{Truzzolillo2015}, is $\varphi = 0.97$ at $T = 20^\circ\mathrm{C}$. The microgel radius (and thus $\varphi$) changes with temperature~\cite{Senff1999}, a property that we will exploit to illustrate the data acquisition scheme for a system with non-stationary dynamics.

\subsection{Microscopy}
Two series of images of colloidal suspensions were taken under an optical microscope (Leica DM IRB), using the variable delay scheme of Sec.~\ref{sec:theo} implemented via the single-thread version of the image acquisition software. The images are taken with a CMOS camera (Basler acA2000-340km, image format $2048\times 1088$ pixels) using a 10x objective, such that one pixel corresponds to $0.55\um$ in the sample. In the first series, we study a suspension of small particles (SP in the following), comprising polystyrene spheres of radius $a= 105$ nm (Microparticles GmbH), diluted to $2.5\times 10^{-3}$ w/w in a 1:1 v/v mixture of $\mathrm{H_2O}$ and $\mathrm{D_2O}$ that matches the density of polystyrene. The second suspension (large particles, LP) contains polystyrene particles with $2a = 1.2\um$ (Invitrogen Molecular Probes), suspended at a weight fraction 0.005\% in the same solvent as the SP. Data for the SP have been analyzed by DDM, while the dynamics of the LP have been quantified by both DDM and particle tracking.

\subsubsection{Differential Dynamic Microscopy}
\label{sec:DDM}
Differential Dynamic Microscopy is a recently introduced technique that combines features of both microscopy and scattering~\cite{Cerbino2008,Cerbino2014}. The dynamics are quantified by a correlation function similar to $g_2(\tau)-1$ introduced above for DLS (see Eq.~\ref{eq:multispeckle}), rather by tracking the motion of individual particles. The analysis is performed on $\tilde{S}(\mathbf{q}, t)$, the Fourier transform of the 2D signal $S(\mathbf{x}, t)$ recorded by the camera, where $\mathbf{x}$ is the coordinate of an image point, and where the $q$ vector has components $q_{x,y} = 2 \pi n_{x,y}/(N_{x,y}l_{p})$, with $0 \leq n_{x,y} \leq N_{x,y}$ and $l_p$ and $N_{x,y}$ the pixel size in the sample and the number of pixels of the field of view along the $x$ and $y$ directions, respectively. For the sake of simplicity and efficiency, the images are cropped to a square format $N_{x} = N_{y} = 1024$ pixels. The quantity of interest is
\begin{equation}
\label{eq:classicDDM}
c_{DDM}(q, t, \tau) = 1 - \frac{\left<\left|\tilde{S}(\mathbf{q}, t)-\tilde{S}(\mathbf{q},t+ \tau)\right|^2\right>_{\mathbf{q}}}{\left<\left|\tilde{S}(\mathbf{q}, t)\right|^2\right>_{\mathbf{q}}+\left<\left|\tilde{S}(\mathbf{q}, t+\tau)\right|^2\right>_{\mathbf{q}}} \,,
\end{equation}
with $\left < \cdot \cdot \cdot \right >$ an azimuthal average over $\mathbf{q}$ vectors with the same magnitude.
Equation~(\ref{eq:classicDDM}) is the degree of correlation corresponding to the structure function normally used in DDM~\cite{Cerbino2014}, except for the normalization factor. Note that while the contribution of static optical noise (due e.g. to dust on the microscope optics or the CMOS sensor) cancels out in the numerator of the last term of the r.h..s of \ref{eq:classicDDM}, it does not vanish in the denominator. As a consequence, the degree of correlation does not fully decay to 0 at large $\tau$, when the scatterers' configuration is completely renewed, but rather to a finite baseline. The optical noise varies with $q$; accordingly, the baseline amplitude is $q$ dependent. It is smallest ($\approx 10^{-2}$) at intermediate $q$ vectors and increases both at larger $q$ (up to a level of $\approx 0.3$) and at small $q$, reaching 0.9999 at the smallest probed scattering vectors. In the following, when presenting DDM data we subtract off the baseline and renormalize the correlation function such that $c_{DDM}(\tau \rightarrow 0) = 1$. As for the DLS data, the DDM two-times degree of correlation, Eq.~(\ref{eq:classicDDM}), is averaged over an appropriate time interval to obtain the DDM intensity correlation function
\begin{equation}
\label{eq:g2classicDDM}
g_{2,DDM}(\tau)-1 = \left< c_{DDM}(q, t, \tau) \right>_t^2 \,,
\end{equation}
where the r.h.s. is squared since $c_{DDM}$ corresponds to a field correlation function~\cite{Cerbino2014}, rather than to an intensity correlation function.

\subsubsection{Far-Field Differential Dynamic Microscopy}
\label{sec:ffddm}
The traditional DDM correlation function, Eq.~(\ref{eq:classicDDM}), is sensitive to any global drift of the sample. A collective drift often arises as a consequence of an artifact; for example, for our Brownian samples a spurious drift motion is sometimes observed, most likely due to convection induced by the sample illumination. It is therefore interesting to use a DDM correlation function that is insensitive to drift. A simple choice inspired by light scattering and proposed by Buzzaccaro et al.~\cite{Buzzaccaro2015} is
\begin{equation}
\label{eq:stefanoDDM}
c_{FF-DDM}(q, t, \tau) = \frac{\left<\left|\tilde{S}(\mathbf{q}, t)\tilde{S}(\mathbf{q},t+ \tau)\right|^2\right>_{\mathbf{q}}}{\left<\left|\tilde{S}(\mathbf{q}, t)\right|^2\right>_{\mathbf{q}}\left<\left|\tilde{S}(\mathbf{q}, t+\tau)\right|^2\right>_{\mathbf{q}}} - 1 \,
\end{equation}
and the associated time-averaged function
\begin{equation}
\label{eq:g2stefanoDDM}
g_{2,FF-DDM}(\tau)-1 = \left< c_{FF-DDM}(q, t, \tau) \right>_t \,.
\end{equation}
The subscript FF-DDM stands for far-field DDM, since in Eq.~(\ref{eq:stefanoDDM}) the correlation function is calculated on the square of the Fourier transform of $S$, which corresponds to the far-field intensity distribution that would be observed in a light scattering experiment. The expression above is independent of the phase of $\tilde{S}$: this makes it insensitive to any global drift, except for the decorrelation arising from the fact that, due to drift, some particles may leave the field of view and be replaced by incoming particles, whose position will in general be totally uncorrelated with respect to that of the scatterers leaving the field of view. This loss of correlation is of course negligible if the drift is much smaller with respect to the field of view, which is the case in our experiments. Finally, we note that optical noise affects the FF-DDM degree of correlation, similarly to the case of the quantity introduced in Eq.~\ref{eq:classicDDM}. We therefore subtract off the data the large-$\tau$ baseline, whose amplitude is comparable to that discussed above for DDM .

\subsubsection{Particle tracking}
\label{sec:tracking}
The series of microscope images acquired for the large colloids was also analyzed by tracking the motion of the particles, in order to extract the mean square displacement from real-space measurements. The Python trackpy package~\cite{dan_allan_2014_12255} was used; by applying filters on the particle shape and size, particles out of focus were rejected. All particle trajectories lasting less than 50 s (because the particles left the field of view or the focal plane) were discarded. For each $\tau$, the 2D mean squared displacement is obtained by averaging over at least $N_t \sim 1000$ trajectories:
\begin{equation}
\left<\Delta r^2 (\tau) \right> = \left < N_t^{-1} \sum_{i=1}^{N_t} \left [\Delta x_i^2 (t,\tau) +  \Delta y_i^2 (t,\tau) \right] \right >_t \,,
\label{eq:msd}
\end{equation}
with $\Delta x_i(t,\tau)$, $\Delta y_i(t,\tau)$ the $x$ and $y$ components of the particle displacement between times $t$ and $t+\tau$ for the $i-$th trajectory. When calculating $<\Delta r^2>$, we reject the contribution of drift motion: for each pairs of frames, the average particle displacement is subtracted off, so that $\sum_{i=1}^{N_t}  \Delta x_i = \sum_{i=1}^{N_t}  \Delta y_i =0$.

\section{Results and discussion}
\label{sec:results}

\begin{figure}[htbp]
\centering{
\includegraphics[width=8cm]{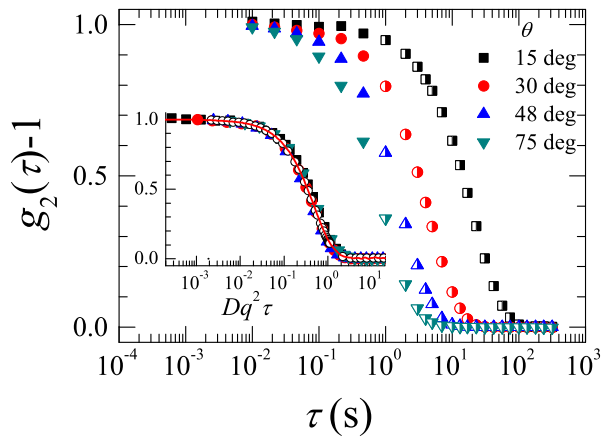}
\caption{Main plot: intensity correlation functions measured simultaneously at four scattering angles $\theta$ by multispekle DLS, for a diluted suspension of melamine particles with diameter $2a=1.14\um$. The solid symbols are the sub-$\tpp$ delays, the semi-filled symbols are integer multiples of $\tpp$. For the latter, $g_2-1$ is plotted only for selected delays, to avoid overcrowding the plot. Inset: $g_2-1$ versus time rescaled by $Dq^2$ for the same data as in the main figure (same symbols), and for a diluted suspension of melamine particles with $2a = 1.6\um$ (open circles). The line is $g_2-1$ for the $1.6\um$ particles as measured by conventional DLS, under the same conditions as for the multispeckle experiment. See the text for the details on the acquisition scheme used in the multispeckle measurements.}}
\label{figdlsa}
\end{figure}

Figure~2 
illustrates an application of the variable-delay acquisition scheme to a DLS experiment. The intensity correlation functions have been obtained from data collected simultaneously at scattering angles $\theta = 15, 30, 48,$ and 75 deg, for a diluted suspension of melamine particles. At each angle, 960 images have been acquired using the following parameters: $\tpp = 1~\mathrm{s}$, $\taumin = 10^{-2}~\mathrm{s}$ and $J = 3$ points per decade. The resulting average data acquisition rate is $8\times 10^5~\mathrm{bytes~s^{-1}}$, a factor $\xi = 50$ less than what it would have been by acquiring the images at a constant rate $\taumin^{-1} = 100~\mathrm{Hz}$. The solid symbols correspond to sub-$\tpp$ delays that are obtained from pairs of consecutive even and odd images. The semi-filled symbols correspond to integer multiples of $\tpp$: they are obtained from pairs of even images. [Only some selected time delays multiple of $\tpp$ are shown in Fig.~2]. The inset of Fig.~2 
shows the same data as in the main figure, re-plotted \textit{vs} the scaled time $Dq^2 \tau$. The open circles are additional data for a diluted suspension of melamine particles with $2a = 1.6\um$, also acquired using the variable-delay scheme. All data collapse on a master curve that follows the correlation function measured for the $1.6\um$ melamine beads with a commercial DLS apparatus (Brookhaven AT2000, red line). Since any deviation from the prescribed temporal acquisition scheme would result in an artifactual change of $g_2-1$, the collapse shown in the inset of Fig.~2 
demonstrates that the variable-delay acquisition and buffering scheme works correctly.

\begin{figure}[htbp]
\centering{
\includegraphics[width=8cm]{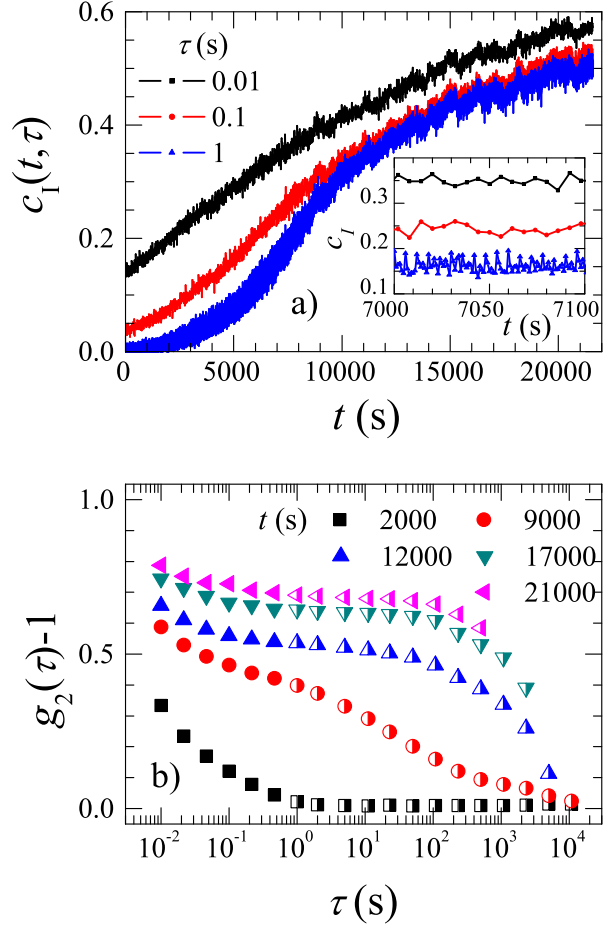}
\caption{a): time evolution of the degree of correlation $c_I$, Eq.~\ref{eq:cI}, for 3 selected delays, as indicated by the label. Data are obtained from multispeckle DLS measurements at $\theta = 46\rmdeg$ for a suspension of PNiPAM microgels whose volume fraction slowly increases during the measurements (see text for details). Inset: zoom of the data showing that the dynamics are stationary over a short $t$ interval. b): $g_2-1$ obtained by averaging $c_I(t,\tau)$ over a time window of 100 s, for various starting times after the beginning of the experiment, as indicated by the label. Filled and semi-solid symbols correspond to sub-$\tpp$ and integer multiples of $\tpp$, respectively. For the latter, $g_2-1$ is plotted only for selected delays. }}
\label{figdlsb}
\end{figure}

One advantage of the scheme proposed in this work is to cover a wide range of delay times without alternating between series of images taken at a fast and slow rate, which makes it suitable for system whose dynamics evolve in time. An example is given in Fig.~3, 
where we show data obtained by multispeckle DLS for a suspension of thermosensitive PNiPAM microgels. The acquisition parameters used in this experiment are the same as for those in Fig.~2. Figure~3a shows the time evolution of the degree of correlation $c_I(t,\tau)$ for three time delays $\tau$, as shown by the label. As a general trend, $c_I$ increases with time, a behavior typical of systems whose dynamics slow down~\cite{Duri2005}. Here, the slowing down of the dynamics is due to a change of the volume fraction: throughout the experiment, $T$
decreases at a constant rate $\dot{T} \approx 3.7 \times 10^{-4}~^\circ\mathrm{C~s^{-1}}$, which results in a growth of the microgel size and thus of their volume fraction, from $\varphi \approx 0.65$ at $t = 0$ to $\varphi \approx 0.97$ at $t = 21000~\mathrm{s}$. Thanks to the variable delay scheme, it is possible to follow the evolution of the dynamics with a good temporal resolution: for the two sub-$\tpp$ delays shown in Fig.~3a, $c_I(t,\tau)$ can be calculated once per cycle (every 6 s), while for $\tau = 1~\mathrm{s}$ data are available every $\tpp = 1~\mathrm{s}$. Such detailed information is useful since any local deviation with respect to the general trend may reveal an experimental problem, or simply because the rate of change of the dynamics may not be known beforehand, thus making it impossible to optimize the acquisition parameters \textit{a priori}. Detailed knowledge of the time evolution of $c_I$ provides also guidance for choosing the time window $t_{exp}$ over which the data may be averaged in order to calculate $g_2-1$. Figure 3a shows that the growth of $c_I$ is steepest around $t = 7000~\mathrm{s}$. Accordingly, $t_{exp}$ should be small enough for $c_I$ not to change significantly in the worst-case scenario, i.e. for $\tau = 1~\mathrm{s}$ and around $t = 7000~\mathrm{s}$. The inset of Fig.~3a shows that $t_{exp} = 100~\mathrm{s}$ satisfies this criterion: we therefore average $c_I$ over such a time window in order to reduce the experimental noise without loosing information on the evolution of the dynamics. Figure 3b shows $g_2-1$ thus obtained for selected values of $t$. As $t$ grows, the volume fraction of the suspension increases due to the swelling of the microgels and the decay of $g_2-1$ shifts to longer times, while the shape of the correlation function changes from a single mode relaxation to a two-step decay, a behavior typical of dense colloidal suspensions~\cite{Hunter2011}. These changes are very well captured by using the variable-delay scheme, which allows to measure efficiently $g_2-1$ over 6 orders of magnitude in $\tau$.

\begin{figure}[htbp]
\centering{
\includegraphics[width=8cm]{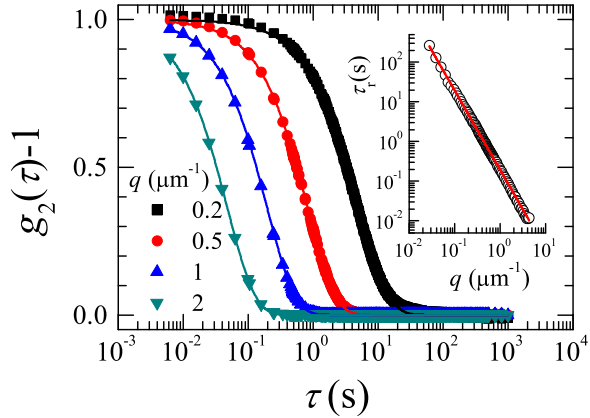}
\caption{Representative correlation functions obtained by differential dynamic microscopy for the SP sample. The data (symbols) are labelled by the corresponding scattering vector, the lines are exponential fits to the decay of $g_2-1$. The amplitude of the baseline that has been subtracted (from the smallest to the largest $q$) is 0.848, 0.037, 0.015, 0.071. Inset: relaxation time extracted from the fits, as a function of $q$. The line is a power law fit to the data, yielding an exponent $-2.01 \pm 0.01$.}}
\label{figmicroscopea}
\end{figure}

As an example of the variable-delay scheme applied to microscopy experiment, Fig.~4 
shows representative correlation functions obtained by conventional DDM for the SP sample. The experimental parameters are $\tpp = 0.5~\mathrm{s}$, $\taumin = 4\times 10^{-3}~\mathrm{s}$ and $J = 5$ points per decade, corresponding to a compression factor $\xi = 62.5$. The data are very well fitted by an exponential decay, $g_2(\tau)-1 = \exp(-\tau/\tau_r)$ (lines). The inset shows the relaxation time $\tau_r$ extracted from the fit as a function of $q$ vector. The line is a power law fit to $\tau_r(q)$, yielding an exponent $-2.01 \pm 0.01$, fully consistent with the $q^{-2}$ scaling expected for a diffusive process~\cite{Berne1976}. Both the shape of $g_2-1$ and the $q$ dependence of the relaxation time are in excellent agreement with those expected for Brownian particles: this demonstrates that the variable-delay scheme works correctly and that the simple single-thread implementation used here is a viable alternative to a more complex multi-thread acquisition software.

\begin{figure}[htbp]
\centering{
\includegraphics[width=8cm]{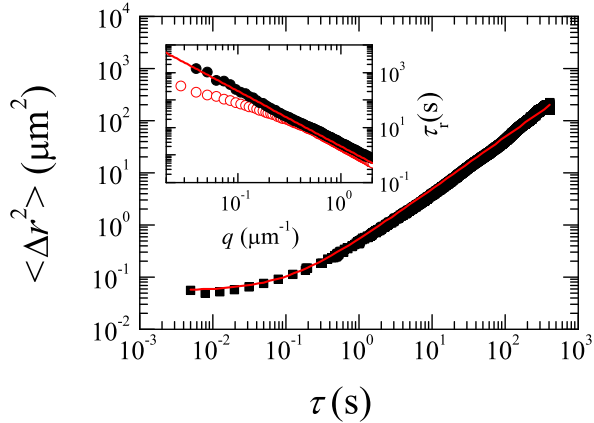}
\caption{Mean squared displacement for the LP sample, calculated from the particle trajectories obtained by video microscopy. The deviation from a linear behavior at small $\tau$ is due to the noise of the tracking algorithm. Inset: Relaxation time vs $q$ issued from a conventional DDM (open symbols) or far field-DDM analysis of the same series of images as in the main panel. The line is the expected behavior for Brownian particles with the same $D$ as in the main panel.}}
\label{figmicroscopeb}
\end{figure}

For the SP sample, the particles are too small to be directly visualized by microscopy: accordingly, direct space techniques cannot be applied to them. By contrast, the data for the LP sample can be analyzed both by tracking the particle trajectories and by DDM. The main plot of Fig.~5 
shows $<\Delta r^2>$ obtained by tracking the particles in a series of images taken with the variable-delay method, with parameters $\tpp = 0.5~\mathrm{s}$, $\taumin = 5\times 10^{-3}~\mathrm{s}$ and $J = 5$ points per decade. At large $\tau$, $<\Delta r^2>$ scales with $\tau$, as expected for Brownian motion, whereas at low $\tau$ the mean square displacement tends to a constant value. This behavior is due to the uncertainty in the particle position as determined by the tracking algorithm~\cite{Crocker1996}. To account for the tracking errors, we fit the data with the affine law $<\Delta r^2> = 4Dt + 4\varepsilon^2$, where $\varepsilon$ is the rms tracking error on each coordinate and the first term on the r.h.s. accounts for 2D diffusive motion. As shown by the red line, the data are very well fitted by this expression, with $D = 0.122\um^2\mathrm{s}^{-1}$ and and error $\varepsilon = 0.12~\um$ (corresponding to 0.2 pixel) comparable to that typically achievable by low-magnification optical microscopy~\cite{Elliot2001}. The inset of Fig.~5 
shows the results of a DDM analysis of the same series of images. At large $q$, the relaxation time obtained from an exponential fit of the conventional DDM correlation function, (Eqs.~\ref{eq:classicDDM},\ref{eq:g2classicDDM}, open red circles), follows the expected $q^{-2}$ scaling. However, at small $q$ $\tau_r$ strongly deviates from this behavior, since the relaxation time is increasingly lower than expected as $q$ decreases. A plausible explanation of these observations is that the particles undergo collective drift motion, in addition to Brownian diffusion. A possible source of drift is convective motion triggered by heating due to sample illumination. Note that collective drift is corrected for by the particle tracking algorithm (see Sec.~\ref{sec:tracking}). This explains why no deviations from diffusive motion are observed in the main plot at large values of $<\Delta r^2>$, which correspond to the small $q$ regime of the inset. We apply the FF-DDM algorithm, Eqs.~\ref{eq:stefanoDDM},\ref{eq:g2stefanoDDM}, to the same series of images: the relaxation time thus obtained (solid black circles in the inset of Fig.~5 
follows the expected diffusive behavior, with no roll-off at small $q$. Moreover, the data are in good agreement with the red line, which shows the behavior expected for diffusive motion with the same diffusion coefficient as that obtained from the fit of $<\Delta r^2>$.
We thus conclude that the variable-delay scheme once again works correctly and that the far-field DDM method is effective in suppressing spurious contributions due to a global drift of the particles.

\section{Conclusions}
\label{sec:conclusions}
We have introduced a variable-delay temporal scheme that allows data to be acquired at a low average rate, while still sampling the dynamics over a wide range of characteristic times, including times much shorter than the inverse average acquisition rate. This scheme has been demonstrated in light scattering and microscopy experiments on colloidal suspensions, where the setups comprise one or more CCD or CMOS cameras that generate large data flows. In analyzing the microscopy data, we have validated far-field DDM, a variant~\cite{Buzzaccaro2015} of the recently introduced DDM method, which allows one to reject the contribution of a global drift to the measured dynamics, e.g. as due to convective motion, slight sample evaporation, or setup vibrations.

Since the acquisition scheme proposed in this paper undersamples the system, it leads in principle to poorer average than that theoretically achievable if data were acquired at the maximum rate. However, this loss of information is more than offset by the ease of coping with a reduced average data flow rate. This is a valuable feature when large amounts of data are generated, as for the 2D detectors in our DLS and microscopy experiments. Another potential application is the processing of relatively small data streams, but with low-cost, low-performance hardware, e.g. based on an Arduino card and a single-board computer such as the Rasberry Pi, or a mobile app run on a smartphone. Setups based on similar hardware are now seen as a valuable alternative to more costly, traditional instruments, e.g. for educational purposes or for developing countries~\cite{Kelley2014}.\newline

\textit{Acknowledgments}. The research leading to these results has
received funding from CNES, the French ANR (project FAPRES) and from the European Research Council
under the European Union’s Seventh Framework
Programme (FP7/2007-2013)/ERC Grant Agreement
No. 306845. We thank Y. Nagazi and E. Tamborini for help in the experiments, and R. Cerbino, M. Alaimo, and F. Giavazzi for enlightening discussions on DDM.\newline

\textbf{Bibliography}\\
\bibliographystyle{unsrt}
\bibliography{smart_sampling2}

\newpage

\section*{Supporting Data: demonstration Python code}

We briefly describe here the Python code reported below and used to demonstrate the multiple-delay acquisition and processing scheme in its single-thread implementation. The code is also available as an uncommented, stand-alone file, see Supplementary data.

Lines 1-133 contain various functions used to generate the acquisition times, implement the acquisition, analyze the data. Lines 141-255 are the main, which demonstrates simple data acquisition and processing using the above functions. The user may edit the input parameters that control the acquisition and process, lines 147-178.

The main routine first demonstrates how to calculate the various quantities that define the acquisition scheme (lines 182-200), in particular the scheduled acquisition times for all images (line 186, call to \verb|acquisition_time()|).

Next, we demonstrate the use of the function \verb|find_pairs()| (lines 203-212). This is a utility function that returns a list of all pairs of images separated by a given time delay $\tau$; \verb|find_pairs()| is typically used in order to extract the appropriate pairs of images when, e.g., calculating an intensity correlation function or a particle mean square displacement for a given time delay.

Lines 214-244 illustrate how to implement the multiple-delay acquisition and processing scheme. For the sake of simplicity, no external acquisition device is used here: when the \verb|acquire_data()| function is called within the acquisition loop (line 233), a counter is incremented and stored in the acquisition buffer. At each iteration of the acquisition loop, the function \verb|can_process()| is called to check whether processing can be performed (line 266). Processing is done only if enough time is available before the next scheduled acquisition and if the acquisition buffer has not already been emptied. In this demo, processing is performed within the \verb|process_data()| function (called at line 228) and simply consist of writing an 8-bit image to the hard disk, with all pixels set to an intensity level given by the acquisition counter modulus 255. At the end of the acquisition loop, the acquisition buffer is emptied by processing all data that were not yet processed (lines 237-242). Finally, in lines 246-255 some quantities are calculated for checking purposes and printed to the screen.\\
\\

\begin{lstlisting}
#! /usr/local/bin/python
# -*- coding: utf-8 -*-
# Demonstration code for the multiple-delay
# acquisitiàon scheme described in
# 'An efficient scheme for sampling fast dynamics at a low average
# data acquisition rate', by A. Philippe et al., submitted to
# J. Phys.:Cond .Mat.

def acquisition_time(tau_min,J,t_pp,M):
    '''generates a list of acquisition times over M cycles
       returns the list asa numpy array'''

    from math import ceil,log10
    import numpy as np
    time_acq= []
    N = int(ceil(J*log10(t_pp/tau_min)))
    cycle_dur = N*t_pp
    for cycle in range(1,M+1):
        for image in range(0,2*N):
            if (image % 2) == 0:
                tt = (cycle-1)*cycle_dur + image/2 * t_pp
            else:
                tt = (cycle-1)*cycle_dur + (image-1)/2 * t_pp + \
                     10**(float(image-1)/(2*J))*tau_min
            time_acq.append(tt)
    return N, np.asarray(time_acq)

def wait_untill(tt,t0):
    '''waits untill a call to time.clock() returns t0+tt,
       where t0 is a reference starting time and tt is a time delay
       returns the actual delay elapsed since t0'''

    from time import clock
    while True:
        elapsed = clock()-t0
        if elapsed >= tt: break
    return elapsed

def process_data(buffer_data,n_processed,output_folder):
    '''as an example of image 'processing', here we write to the hard disk a 8-bit
    image of size 640x480 pixels. The intensity level is set to the image number,
    modulus 255.
    Requires the Python Image Library (PIL)'''

    from PIL import Image
    buf_size = buffer_data.size
    pix_val = buffer_data[(n_processed+1)%buf_size]
    im = Image.new("L",(640,480),int(pix_val))
    im.save(output_folder+str(n_processed+1).zfill(3)+'.jpg')
    return

def acquire_data(t0,time_acq,n_acquired,buffer_data):
    '''routine to acquire data and store them in a buffer (e.g. grab an image)
       It returns the time (with respect to t0) when the data were actually acquired
       In this example, we simply write one single 8-bit unsigned integer to the buffer
       '''

    nacq = n_acquired+1 # the nacq-th data will be acquired
    buf_size = buffer_data.size
    tt = time_acq[nacq]
    elapsed = wait_untill(tt,t0)  # wait the correct amount of time before acquiring data
    data_value = nacq%255  # here the data is acquired
    buffer_data[nacq%buf_size] = data_value #store the data in the buffer
    return elapsed

def can_process(t0,time_acq,process_time,n_acquired,n_processed):
    '''check if data can be processed. Two conditions must be fullfilled:
       1) there is enough time left before the next data point acquisition
       2) the buffer has not already been read (including a safety 'lag' of 4
        data points)'''

    from time import clock
    time_to_acq = time_acq[n_acquired+1]-(clock()-t0) #time to next data acquisition
    condition = (time_to_acq > process_time) and (n_acquired > n_processed + 4)
    # note: leave a safety 'lag' of 4 data points between acquired data and processed data
    if (condition): # there is enough time to process data
        return True
    else:
        return False


def find_pairs(tau,time_acq,tw1,tw2):
    '''given the numpy vector of the acquisition times time_acq (typically generated
       using acquisition_time(tau_min,J,t_pp,M) ), finds all pairs of images separated
       by the time delay closest to the desired tau and such that the
       first image of each pair was acquired at a time tw1 <= t <= tw2
       Returns:
       image1, image2: numpy vectors with the index of the first and second
                       images of all acceptable pairs, respectively
       actualdelay: the available delay that best matches the required input delay tau
       tw1, tw2: the limits of the t range, updated and corrected if those given
                  as an input where not acceptable'''

    from math import modf
    rtol = 1E-5    #relative tolerance when testing the delay
                   #that best matches the target one
    #find time per pair t_pp and im_per_cycle, the number of images per cycle
    t_pp = time_acq[2]-time_acq[0]
    tau_min = time_acq[1]-time_acq[0]
    lta = time_acq.size
    for i in np.arange(2,lta-1,2):
        if np.isclose((time_acq[i+1]-time_acq[i]),tau_min):
            im_per_cycle = i
            break
    tw1 = abs(tw1); tw2 = abs(tw2)
    tw1 = min(tw1,time_acq[lta-1])
    tw2 = min(tw2,time_acq[lta-1])
    tw2 = max(tw1,tw2)
    image1_first = np.where(np.asarray(time_acq)>=tw1)[0][0]
    itemp = np.where(np.asarray(time_acq)>tw2)[0]
    if itemp.size>0: image1_last = itemp[0]
    else: image1_last = lta-1

    olddelta = time_acq[lta-1]
    for i in reversed(range(image1_first,image1_last+1)):
        for j in range(i,lta):
            newdelta = abs((time_acq[j]-time_acq[i])-tau)
            if newdelta <= olddelta*(1.+rtol):
                olddelta = newdelta
                igood = i
                jgood = j
    deltai = jgood-igood
    actualdelay = time_acq[jgood]-time_acq[igood]
    mdf = modf(actualdelay/t_pp)[0]
    if np.isclose(mdf,0.): #delay is a multiple of t_pp
       step = 2
    else: step = im_per_cycle
    image1 = np.arange(igood,image1_last+1,step)
    image2 = image1 + deltai
    whereok = np.where(image2<lta)
    image1 = image1[whereok]
    image2 = image2[whereok]
    return image1,image2,actualdelay,tw1,tw2




##############################################################
################# the main ###################################
##############################################################
if __name__ == '__main__':

    from time import clock
    import numpy as np
    import sys as sy

#########################################################################
# edit the following parameters as needed
#########################################################################
# parameters that define the acquisition cycle
    tau_min = 0.015   # minimum delay, sec
    J = 3             # number of sub-t_pp delays per decade
    t_pp = 1.          # time per pair of images, sec
    M = 10            # number of cycles
#
###############################################################
#  parameters to test the function find_pairs()
    target_delay = 1.57
    tw1 = -11.2
    tw2 = 2000
#
###############################################################
# parameters that define the data to be acquired and processed
# for testing purposes.
# The 'acquired' data will be simply the sequence na = 0, 1, 2 ...
# 'processing' will consist in writing to disk an image whose pixels are
#  all set to the value na
    buffer_size = 10  # number of acquired data that can be stored in a buffer
    #create a buffer where to store the acquired data:
    buffer_data = np.zeros(buffer_size) + 0.1
    output_folder = 'c:/temp/_test_buffering/' #acquired data will be written here
    process_time = 0.2   # time to process a data point, sec. (e.g time to write an image
                         # to disk). Note that no processing will be done if the time to
                         # the next data acquisition is less than process_time.
                         # Here we use an exaggerated value for demo purposes
#
# end of parameters to be edited
###############################################################



#### Calculate various quantities that define the acquisition scheme
####
    print '************ Acquisition scheme parameters ********'
    #generate list of acquisition times:
    N, time_acq = acquisition_time(tau_min,J,t_pp,M)
    #total number of data to be acquired and processed:
    tot_num_data = M*2*N
    actual_time = np.zeros(tot_num_data)  #storage space for actual acquisition time

    print 'Number of cycles to be performed:', M
    print 'Images per cycle: ',2*N
    print 'Duration of one cycle:',N*t_pp, 'sec'
    print 'Total experiment duration (sec):',time_acq[tot_num_data-1]
    print 'Acquisition times (sec, only for ther first cycle):\n',time_acq[:2*N]
    #demonstrate how to access the time delay between any pair of images
    i1 = tot_num_data/11; i2 = tot_num_data/2
    print '\nimage',i1,'will be taken at time',time_acq[i1],'sec'
    print 'image',i2,'will be taken at time',time_acq[i2],'sec'
    print 'lag between images',i1,'and',i2,'will be ',time_acq[i2]-time_acq[i1],'sec'


#### Test the function find_pairs()
####
    print '\n************ Test of find_pairs() ********'
    image1,image2, actualdelay, tw1, tw2 = find_pairs(target_delay,time_acq,tw1,tw2)
    print 'tw1,tw2 (may have been corrected)',tw1,tw2
    print 'target delay, actual delay',target_delay,actualdelay
    print 'image1',image1, '\ntime_acq[image1]',time_acq[image1]
    print 'image2',image2, '\ntime_acq[image2]',time_acq[image2]
    print 'image2-image1',image2-image1
    print 'time_acq[image2]-time_acq[image1]:',time_acq[image2]-time_acq[image1]

#### Real-time test of acquisition/data storage
####
    print '\n************ Real-time test of acquisition/data storage ********'
    print '\nNow starting acquisition...'; sy.stdout.flush()
    #initialize variables
    n_acquired = -1   # id of the last acq. data (n_acquired = 0 for the first data)
    n_processed = -1  # id of the last proc. data (n_processed = 0 for the first data)

    t0 = clock()  # get starting time
    for tt in time_acq: #loop over all acquisition times
        # process data, if the buffer needs to be emptied and if enough time is available
        while True:
            do_proc =  can_process(t0,time_acq,process_time,n_acquired,n_processed)
            if do_proc:
                process_data(buffer_data,n_processed,output_folder)
                n_processed = n_processed + 1
            else: break

        #acquire data
        actual_time[n_acquired+1] = acquire_data(t0,time_acq,n_acquired,buffer_data)
        n_acquired = n_acquired + 1
    # end of data acquisition loop

    # process the acquired data that are still in the buffer but
    # have not being processed yet
    print 'n_processed,n_acquired', n_processed,n_acquired
    print 'Now processing the last',n_processed,n_acquired,'data'
    for nproc in range(n_processed,n_acquired):
        process_data(buffer_data,nproc,output_folder)

    print 'done!\n\n'

    #check if the acquisition times were correct, print some info
    delta = actual_time-time_acq
    print 'actual acquisition time - scheduled acquisition time (sec):'
    print delta
    print 'Max difference between actual and scheduled acquisition time (sec):'
    print delta.max()
    print 'Min difference between actual and scheduled acquisition time (sec):'
    print delta.min()
    print 'Rms difference between actual and scheduled acquisition time (sec):'
    print delta.std()
\end{lstlisting}


\end{document}